\documentclass[aps,preprint,showkeys,showpacs]{revtex4}
\input{epsf}

\begin{document}
\newcommand{\be}{\begin{equation}}
\newcommand{\ee}{\end{equation}}
\newcommand{\bea}{\begin{eqnarray}}
\newcommand{\eea}{\end{eqnarray}}

\title{Partition function zeros of the $Q$-state Potts model on the simple-cubic lattice}
\author{Seung-Yeon Kim\footnote{Electronic address: sykim@kias.re.kr}}
\affiliation{School of Computational Sciences,
Korea Institute for Advanced Study,\\
207-43 Cheongryangri-dong, Dongdaemun-gu, Seoul 130-012, Korea}

%\date{\today}

\begin{abstract}
The $Q$-state Potts model on the simple-cubic lattice is studied
using the zeros of the exact partition function on a finite lattice.
The critical behavior of the model in the ferromagnetic and
antiferromagnetic phases is discussed based on the distribution
of the zeros in the complex temperature plane.
The characteristic exponents at complex-temperature singularities,
which coexist with the physical critical points in the complex temperature plane
for no magnetic field ($H_q=0$), are estimated using the low-temperature series expansion.
We also study the partition function zeros of the Potts model
for nonzero magnetic field. For $H_q>0$ the physical critical points disappear and
the Fisher edge singularities appear in the complex temperature plane.
The characteristic exponents at the Fisher edge singularities are calculated
using the high-field, low-temperature series expansion.
It seems that the Fisher edge singularity is related to the Yang-Lee edge singularity
which appears in the complex magnetic-field plane for $T>T_c$.
\end{abstract}

\pacs{05.50.+q; 64.60.Cn; 75.10.Hk; 11.15.Ha}
\keywords{partition function zeros; ferromagnetic transition;
antiferromagnetic phases; complex-temperature singularity; Fisher edge singularity;
Yang-Lee edge singularity}

\maketitle

%%%%%%%%%%%%%%%%%%%%%%%%%%%%%%%%%%%%%%%%%%%%%%%%%%%%%%%%%%%%%%%%%%%%%%%%%%%

\section{introduction}

The $Q$-state Potts model\cite{wu} in two and three dimensions
exhibits a rich variety of critical behavior
and is very fertile ground for the analytical and numerical
investigation of first- and second-order phase transitions.
With the exception of the two-dimensional $Q=2$ Potts (Ising) model
in the absence of an external magnetic field,
exact solutions for arbitrary $Q$ are not known.
However, some exact results at the critical temperature have been established for the
two-dimensional $Q$-state Potts model for no magnetic field.
From the duality relation the ferromagnetic critical temperature
is known to be $T_c=J/k_B{\rm ln}(1+\sqrt{Q})$ for the isotropic square lattice.
The ferromagnetic Potts model has a first-order phase transition for $Q>4$ in two
dimensions\cite{baxter73}, and has been the most important means to understand
first-order transitions which are omnipresent in nature, not only in condensed-matter
physics but also in subatomic physics, for example, the deconfining or chiral
symmetry breaking transition in quantum chromodynamics\cite{brown,alves90a}
and phase transitions in the very early Universe\cite{kolb}.
In three dimensions the ferromagnetic $Q$-state Potts model for
$Q>Q_c\approx2.45(1)$\cite{jlee} has a first-order transition,
and the three-state Potts model is related to the order parameter (the expectation
value of the Wilson line) for the deconfinement transition
in quantum chromodynamics\cite{svetitsky}.

On the other hand, the antiferromagnetic Potts model is much less well
understood than the ferromagnetic model.
One of the most interesting properties of the antiferromagnetic model is that
for $Q>2$ the ground-state is highly degenerate and the ground-state entropy is nonzero.
The Baxter conjecture\cite{baxter82} for the critical temperature of the Potts
antiferromagnet ($J<0$) on the square lattice, $T_c=J/k_B{\rm ln}(\sqrt{4-Q}-1)$, gives
the known exact value for $Q=2$, a critical point at zero temperature for $Q=3$,
and no critical point for $Q>3$.
In three dimensions the Potts model on the simple-cubic lattice has the
antiferromagnetic ordered phase for $Q\le4$\cite{banavar}.
It is now commonly accepted that the order-disorder phase transition of
the three-state Potts antiferromagnet on the simple-cubic lattice belongs to
the universality class of the three-dimensional $XY$ model\cite{banavar,wang}.
Recently Rosengren {\it et al.}\cite{rosengren93,kundrotas} claimed the existence of
two additional phase transitions below the order-disorder phase transition for
the simple-cubic three-state Potts antiferromagnet.
However, Kolesik and Suzuki\cite{kolesik}, Heilmann {\it et al.}\cite{heilmann},
and Oshikawa\cite{oshikawa} reported different results in disagreement
with observations by Rosengren's group.

By introducing the concept of the zeros of the partition function
in the {\it complex} magnetic-field plane,
Yang and Lee\cite{yang} proposed a mechanism
for the occurrence of phase transitions in the thermodynamic limit
and yielded a new insight into the unsolved problem of the Ising model
in an arbitrary nonzero external magnetic field.
It has been shown\cite{yang,tdlee,creswick97,creswick98,creswick99,janke01}
that the distribution of the zeros of a model
determines its critical behavior. Lee and Yang\cite{tdlee} also
formulated the celebrated circle theorem which states that
the partition function zeros of the Ising ferromagnet lie on the unit circle
in the complex magnetic-field ($X=e^{\beta H}$) plane.
In three dimensions the properties of the zeros in the complex $X$ plane of the
Ising model have been studied by exact enumeration\cite{suzuki} and
series expansion\cite{kortman}.
However, for the $Q$-state Potts model with $Q>2$ the partition function zeros
in the complex $X$ plane lie close to, but not on, the unit circle with the two
exceptions of the critical point $X=1$ ($H=0$) itself and the
zeros in the limit $T=0$\cite{kim98a,kim99,kim00a}.

Fisher\cite{fisher65} emphasized that the partition function zeros in the complex
temperature ($Y=e^{-\beta J}$) plane are also very useful in understanding
phase transitions, and showed that for the square lattice Ising model in the absence of
an external magnetic field the zeros in the complex $Y$ plane lie on
two circles (the ferromagnetic circle $Y_{FM}=-1+\sqrt{2}e^{i\theta}$
and the antiferromagnetic circle $Y_{AFM}=1+\sqrt{2}e^{i\theta}$)
in the thermodynamic limit.
In particular, using the Fisher zeros both the ferromagnetic
phase and the antiferromagnetic phase can be considered at the same time.
Fisher also showed that the logarithmically infinite specific heat
singularity of the Ising model results from the properties of the
density of zeros.
The zeros in the complex $Y$ plane of the simple-cubic Ising model have been studied
extensively to understand the analytical properties and critical behavior
of the model by exact enumeration\cite{ono,pearson,itzykson,martin83a,bhanot90a} and
to determine the critical point and exponents
by Monte Carlo simulation\cite{marinari,bhanot87,alves90b,alves00,janke01}.
Recently it has been shown that for self-dual boundary conditions
near the ferromagnetic critical point $Y_c=1/(1+\sqrt{Q})$
the zeros of the $Q$-state Potts model on a finite square lattice
in the absence of an external magnetic field lie on the circle with center $-1/(Q-1)$
and radius $\sqrt{Q}/(Q-1)$ in the complex $Y$ plane, while the antiferromagnetic circle
of the Ising model completely disappears for
$Q>2$\cite{martin86,cnchen,kim98b,kim00b,kim01}.
Using the distribution of the zeros in the complex temperature plane of the square-lattice
Potts model, the critical behavior in both the ferromagnetic and antiferromagnetic phases
has been studied\cite{kim98b,kim00b,kim01},
and the unknown properties of the model for nonzero external magnetic field have been
invetigated\cite{kim98b}. The partition function zeros in the complex $Y$ plane
of the three-state Potts model on the simple-cubic lattice have been studied
to understand the ferromagnetic critical behavior by exact enumeration
for the $3\times3\times9$ simple-cubic lattice\cite{martin83b}
and by Monte Carlo simulation\cite{alves91,janke01}.

In this paper we study the partition function zeros of the $Q$-state Potts model
for $2\le Q\le7$ on the simple-cubic lattice to unveil some of the rich structures
of the model. The partition function zeros of the three-dimensional Potts model
have never been obtained for $Q>3$ in the literature.
In the next section we describe briefly the microcanonical transfer matrix
to evaluate the density of states, from which the exact partition function
of the simple-cubic Potts model is obtained. In Section 3 we discuss
the ferromagnetic and antiferromagnetic phase transitions and the complex-temperature
singularity (nonphysical critical point) of the simple-cubic Potts model using
the partition function zeros in the absence of an external magnetic field.
We also discuss the properties of the complex-temperature singularity
using the low-temperature series expansion.
In section 4 we study the unknown properties of the simple-cubic Potts model
for nonzero magnetic field using the partition function zeros
and the high-field, low-temperature series expansion.
We discuss the Yang-Lee edge singularity and the Fisher edge singularity,
which are nonphysical critical points for nonzero magnetic field,
and relationship between the Yang-Lee and Fisher edge singularities.

%%%%%%%%%%%%%%%%%%%%%%%%%%%%%%%%%%%%%%%%%%%%%%%%%%%%%%%%%%%%%%%%%%%%%%%

\section{density of states}

The $Q$-state Potts model in an external magnetic field $H_q$ on a
lattice $G$ with $N_s$ sites and $N_b$ bonds is defined by the Hamiltonian
\be
{\cal H}_Q=-J\sum_{\langle i,j\rangle}
\delta(\sigma_i,\sigma_j)-H_q\sum_{k=1}^{N_s}\delta(\sigma_k,q),
\ee
where $J$ is the coupling constant (ferromagnetic model for $J>0$
and antiferromagnetic model for $J<0$), $\langle i,j\rangle$
indicates a sum over nearest-neighbor pairs, $\delta$ is the
Kronecker delta, $\sigma_i=1,2,...,Q$, and $q$ is a fixed integer
between 1 and $Q$. The partition function of the model is
\be
Z_Q=\sum_{\{ \sigma_n \}} e^{-\beta{\cal H}_Q},
\ee
where $\{\sigma_n\}$ denotes a sum over $Q^{N_s}$ possible spin
configurations and $\beta=(k_B T)^{-1}$. If we define the
restricted density of states with energy $0\le E\le N_b$ and
magnetization $0\le M\le N_s$ by
\be
\Omega_Q(E,M)=\sum_{\{\sigma_n \} } \delta(E-\sum_{\langle i,j\rangle}
[1-\delta(\sigma_i,\sigma_j)])\delta(M-\sum_k\delta(\sigma_k,q)),
\ee
which takes on only integer values, then the partition function can be written as
\be
Z_Q(Y,X)=Y^{-N_b}\sum_{E=0}^{N_b}\sum_{M=0}^{N_s}\Omega_Q(E,M) Y^E
X^M,
\ee
where $Y=e^{-\beta J}$ and $X=e^{\beta H_q}$, and states
with $E=0$ ($E=N_b$) correspond to the ferromagnetic
(antiferromagnetic) ground states. From Eq. (4) it is clear that
$Z_Q(Y,X)$ is simply a polynomial in $Y$ and $X$. If we sum the
restricted density of states over $M$, we obtain the density of states
\be
\Omega_Q(E)=\sum_{M=0}^{N_s}\Omega_Q(E,M),
\ee
whose sum over $E$ is equal to the $Q^{N_s}$ spin configurations,
\be
\sum_{E=0}^{N_b}\Omega_Q(E)=Q^{N_s}.
\ee
In the absence of an external magnetic field,
the partition function of the model is given by
\be
Z_Q(Y)=Y^{-N_b}\sum_{E=0}^{N_b}\Omega_Q(E) Y^{E},
\ee
which is again a polynomial in $Y$.

Here we describe briefly the microcanonical transfer matrix
($\mu$TM)\cite{bhanot90a,bhanot90b,stosic,stodolsky,creswick95,kim01} to obtain
exact integer values for the
density of states $\Omega_{Q}(E)$ of the $Q$-state Potts model on
an $L\times L\times N$ simple-cubic lattice with periodic boundary
conditions in the $x$-$y$ plane (horizontal area of $L\times L$) and free
boundaries in the $z$ direction (height of $N$). First, an array,
$\omega^{(1)}$, which is indexed by the energy $E$ and variables
$\sigma_i$, $1\le i\le L^2$ for the first $x$-$y$ plane of sites
is initialized as
\be
\omega^{(1)}(E;\sigma_1,\sigma_2,...,\sigma_{L\times L})=
\delta(E-\sum_{i=1}^{L\times L}[1-\delta(\sigma_i,\sigma_{i+1})]).
\ee
Now each spin in the plane is traced over in turn,
introducing a new spin variable from the next $x$-$y$ plane,
\be
\tilde{\omega}^{(2)}(E;\sigma'_1,\sigma_2,...,\sigma_{L\times L})=
\sum_{\sigma_1}\omega^{(1)}(E-[1-\delta(\sigma'_1,\sigma_1)];
\sigma_1,\sigma_2,...,\sigma_{L\times L}).
\ee
This procedure is repeated until all the spins in the first plane have been traced
over, leaving a new function of the $L\times L$ spins in the
second plane. The bonds connecting the spins in the second $x$-$y$
plane are then taken into account by shifting the energy,
\be
\omega^{(2)}(E;\sigma'_1,\sigma'_2,...,\sigma'_{L\times L})=
\tilde{\omega}^{(2)}(E-\sum_{i=1}^{L\times L}
[1-\delta(\sigma'_i,\sigma'_{i+1})];
\sigma'_1,\sigma'_2,...,\sigma'_{L\times L}).
\ee
This procedure is then applied to each $x$-$y$ plane in turn until the final
($N$th) plane is reached. The density of states is then given by
\be
\Omega_Q(E)=\sum_{\sigma'_1}\sum_{\sigma'_2}...\sum_{\sigma'_{L\times
L} } \omega^{(N)}(E;\sigma'_1,\sigma'_2,...,\sigma'_{L\times L}).
\ee
The permutation symmetry of the $Q$-state Potts model allows
us to freeze the last spin $\sigma_{L\times L}=1$ of each $x$-$y$
plane. Now we need to consider only $Q^{L\times L-1}$ possible
spin configurations in each plane instead of $Q^{L\times L}$
configurations, and we save a great amount of memory and CPU time.
For example, Table I shows the density of states $\Omega_{Q=7}(E)$ of the
seven-state Potts model on the $3\times 3\times 3$ simple-cubic lattice.

The above procedure can be modified in a straightforward way to
calculate exact integer values for the restricted density of
states $\Omega(E,M)$\cite{stosic,creswick95}. For the first $x$-$y$ plane one has,
instead of Eq. (8),
\be
\omega^{(1)}(E,M;\sigma_1,\sigma_2,...,\sigma_{L\times L})=
\delta(E-\sum_{i=1}^{L\times L} [1-\delta(\sigma_i,\sigma_{i+1})])
\delta(M-\sum_{k=1}^{L\times L}\delta(\sigma_k,q)).
\ee
The spins in the first plane are traced over in sequence just as in Eq. (9)
leading to a new function of the spins in the second $x$-$y$ plane;
\be
\tilde{\omega}^{(2)}(E,M;\sigma'_1,\sigma_2,...,\sigma_{L\times L})=
\sum_{\sigma_1}\omega^{(1)}(E-[1-\delta(\sigma'_1,\sigma_1)],M-\delta(\sigma'_1,q);
\sigma_1,\sigma_2,...,\sigma_{L\times L}).
\ee
Once all the spins in an $x$-$y$ plane have been traced over,
the function for the next plane is completed as in Eq. (10)
\be
\omega^{(2)}(E,M;\sigma'_1,\sigma'_2,...,\sigma'_{L\times L})=
\tilde{\omega}^{(2)}(E-\sum_{i=1}^{L\times L}[1-\delta(\sigma'_i,\sigma'_{i+1})],M;
\sigma'_1,\sigma'_2,...,\sigma'_{L\times L}),
\ee
and the restricted density of states is given by
\be
\Omega_Q(E,M)=\sum_{\sigma'_1}\sum_{\sigma'_2}...\sum_{\sigma'_{L\times L}}
\omega^{(N)}(E,M;\sigma'_1,\sigma'_2,...,\sigma'_{L\times L}).
\ee

%%%%%%%%%%%%%%%%%%%%%%%%%%%%%%%%%%%%%%%%%%%%%%%%%%%%%%%%%%%%%%%%%%%%%%%%%%

\section{partition function zeros in the complex temperature plane}

In this section we discuss the properties of the $Q$-state Potts models on
the simple-cubic lattice using the partition function zeros $\{Y_0\}$ or
$\{A_0\}$ of the models in the complex temperature ($Y=e^{-\beta J}$ or $A=e^{\beta J}$)
plane which are obtained by solving the equation
\be
Z_Q(Y_0)=Y_0^{-N_b}\sum_{E=0}^{N_b}\Omega_Q(E) Y_0^{E}=0
\ee
for the exact partition functions on finite lattices.
Figure 1 shows the partition function zeros in the complex
$Y=e^{-\beta J}$ plane of the $Q$-state Potts models for $Q$=2, 4, and 7 on the
$3\times 3\times 3$ simple-cubic lattice with periodic boundary
conditions in the $x$-$y$ plane and free boundary conditions in the $z$ direction.
In Figure 1, near the positive real axis,
the partition function zeros show one-dimensional distributions for all the Potts models
which become denser as $Q$ increases. Since the value of latent heat
of a system is proportional to the density of zeros in the complex temperature
plane\cite{creswick98}, for $Q\ge3$ the latent heat increases at the transition
temperature as Q increases, and the first-order phase transition becomes stronger.
Moreover, the first zeros of these linear distributions will cut the real axis
and be the ferromagnetic (FM) critical points in the thermodynamic limit.
In Figure 1, as $Q$ increases, the first zeros become closer to the real axis
and to the values of the FM critical temperatures estimated in the literature (Table II).
This fact suggests that the first zero obtained for a smaller size of
simple-cubic lattice approach very closely the real critical point in the limit
$Q\to\infty$ where the mean-field theory is known to be exact\cite{mittag,pearce}.

Even for the three-dimensional Ising model the critical temperature has never
been known exactly except for some conjectures, for example,
the Rosengren conjecture\cite{rosengren1}.
However, for the simple-cubic lattice Hajdukovi\'c
conjectured\cite{hajdukovic} the FM critical temperature
\be
Y_c^{FM}(Q)={1\over(Q+\sqrt{2Q-1})^{1/3}}
\ee
of the $Q$-state Potts model based on two assumptions.
The first assumption is that $Y_c^{FM}(Q)$ reduces to the mean-field result
$Y_c^{FM}=Q^{-1/3}$ in the limit $Q\to\infty$\cite{mittag,pearce}.
And the second assumption is that in $d$ dimensions $Y_c^{FM}(Q)$ is
the solution of an equation
\be
Y^{-d}-c_1(d) Y^{-d/2}-c_2(d)=Q-2,
\ee
where $c_1(d)$ and $c_2(d)$ have to be determined,
for example, $c_1=2$ and $c_2=1$ for $d=2$.
As shown in Table II the Hajdukovi\'c conjecture is close to
numerical estimates in the literature for $2\le Q\le7$.
However, for $Q=1$ the Hajdukovi\'c conjecture, $Y_c^{FM}=0.7937...$,
completely disagrees with the numerical estimate $0.753$\cite{gaunt,stauffer}.
We can guess a simpler form than the Hajdukovi\'c conjecture for the critical temperature,
\be
Y_c^{FM}(Q)={1\over{1/\pi}+Q^{1/3}},
\ee
satisfying the first assumption. Eq. (19) reasonably agrees with
numerical estimates for $1\le Q\le7$ as shown in Table II.
Even though the Hajdukovi\'c conjecture and Eq. (19) are incorrect expressions
for the FM critical temperature of the $Q$-state Potts model on the simple-cubic lattice,
we can infer that the true expression for the critical temperature may be something
between the Hajdukovi\'c conjecture and Eq. (19).

Recently using cluster-variation calculations Rosengren and Lapinskas\cite{rosengren93}
claimed the existence of two additional phase transitions
at $T_{RL1}=0.787(-J/k_B)$ and $T_{RL2}=0.7715(-J/k_B)$ below the well-known
antiferromagnetic (AFM) critical temperature $T_c^{AFM}=1.2259(7)(-J/k_B)$\cite{wang}
for the three-state AFM Potts model ($J<0$) on the simple-cubic lattice.
Using the Monte Carlo method Kundrotas, Rosengren, and Lapinskas\cite{kundrotas}
obtained $T_{KRL}=0.68(-J/k_B)$ and supported their earlier claim. However,
Kolesik and Suzuki\cite{kolesik} and Heilmann {\it et al.}\cite{heilmann}
observed different results by Monte Carlo simulations which are in disagreement
with observations by Rosengren's group, and obtained $T_{KH}=1.0(-J/k_B)$
as the intermediate transition temperature in the region below $T_c^{AFM}$.
Furthermore, based on a renormalization-group picture Oshikawa\cite{oshikawa}
argued that there exists only one transition at $T_c^{AFM}$
for the three-state AFM Potts model on the simple-cubic lattice.
Figure 2 shows the partition function zeros in the complex
$A=e^{\beta J}$ plane of the three-state Potts model on the
$3\times 3\times 30$ simple-cubic lattice with periodic boundary
conditions in the $x$-$y$ plane and free boundary conditions in the $z$ direction.
The interval $0\le A\le1$ ($0\le T\le\infty$) is antiferromagnetic ($J<0$)
while the interval $1\le A\le\infty$ ($\infty\ge T\ge0$) ferromagnetic ($J>0$).
There exists a clear one-dimensional distribution of zeros near
$A_{KRL}=0.2298$, in Figure 2, which favors the existence of additional
phase transitions below $A_c^{AFM}=0.4424$. However, we do not know if
the first zero of the distribution near $A_{KRL}$ cuts the real axis
in the thermodynamic limit, and if there exist the true phase transitions
below $A_c^{AFM}$.

We have also obtained two pairs of the complex-temperature (CT) singularities
(Figure 2) using the low-temperature series\cite{vohwinkel,bhanot,guttmann1}
for the three-state Potts model on the simple-cubic lattice.
Figure 2 exhibits accumulations of
partition function zeros near the CT singularities.
These accumulations imply that some thermodynamic quantities
may diverge at the CT singularities which are not physical.
The critical exponents associated with the CT singularities,
which we call the CT critical exponents,
are defined in the usual way,
\be
C_{CT}^0\sim\biggl(1-{Y\over Y_{CT}}\biggr)^{-\alpha_{CT}},
\ee
\be
m_{CT}^0\sim\biggl(1-{Y\over Y_{CT}}\biggr)^{\beta_{CT}},
\ee
and
\be
\chi_{CT}^0\sim\biggl(1-{Y\over Y_{CT}}\biggr)^{-\gamma_{CT}},
\ee
where $Y_{CT}$ is the location of a CT singularity, and
$C_{CT}^0$, $m_{CT}^0$ and $\chi_{CT}^0$ are the singular parts of the specific heat,
spontaneous magnetization, and susceptibility, respectively.
Here, the superscript $0$ represents no external magnetic field.
Because the density of zeros near a CT singularity is given by\cite{fisher65}
\be
g(Y)\sim\biggl(1-{Y\over Y_{CT}}\biggr)^{1-\alpha_{CT}},
\ee
the density of zeros diverges at the CT singularity if $\alpha_{CT} > 1$.

Guttmann {\it et al.}\cite{guttmann2,guttmann3,guttmann4} studied
the locations of the CT singularities and the CT critical exponents
for the Ising model in detail, introducing a new method of low-temperature
series analysis.
For the Ising model on the simple cubic, body-centered cubic,
and face-centered cubic lattices there exist the CT singularities,
closer to the origin than the physical critical points, which control
the low-temperature series and spoil the analysis of the real critical behavior.
Guttmann {\it et al.}\cite{guttmann2} reported that for the Ising model on
the simple-cubic lattice the CT singularity is located at $(Y_{CT})^2=-0.2860(2)$
($|(Y_{CT})^2|<(Y_c^{FM})^2=0.4120$) with
$\alpha_{CT}\sim1$, $-{1\over8}\le\beta_{CT}\le-{1\over16}$, and
$\gamma_{CT}\sim{9\over8}$.
Itzykson {\it et al.}\cite{itzykson} studied the zeros of the exact partition function
in the complex temperature plane of the Ising model on the $4\times4\times4$
simple-cubic lattice, and found that the distribution of zeros supports
the real existence of the CT singularities.
By the analysis of the longer series they obtained
$\alpha_{CT}=1.25(15)$, $\beta_{CT}=-0.05$, and $\gamma_{CT}=1.1(1)$ for
the Ising model on the simple-cubic lattice, and
argued that the values of the CT critical exponents of the Ising model
are same for the simple cubic, body-centered cubic, and face-centered cubic lattices
and satisfy $\alpha_{CT}+2\beta_{CT}+\gamma_{CT}\simeq2$. Finally,
Itzykson {\it et al.} conjectured that the behavior at the CT singularities is universal.
Recently Guttmann and Enting\cite{guttmann5} obtained $(Y_{CT})^2=-0.2853(3)$ with
$\alpha_{CT}\approx1.03$, $\beta_{CT}\approx-0.01$, and $\gamma_{CT}\approx1.01$
using the low-temperature series generated by the finite-lattice
method of series expansion for the simple-cubic Ising model.
They proposed $\alpha_{CT}=1$, $\beta_{CT}=0$, and $\gamma_{CT}=1$ with
logarithmic corrections, satisfying the Rushbrooke scaling law
$\alpha_{CT}+2\beta_{CT}+\gamma_{CT}=2$.

Using Dlog Pad\'e approximants\cite{guttmann6}
to the low-temperature series\cite{guttmann1} of the specific heat,
spontaneous magnetization, and susceptibility
for the three-state Potts model on the simple-cubic lattice,
we found the locations of two pairs of the CT singularities
\be
Y_{CT}^{(1)}=0.0254(11)\pm0.4948(9)i
\ee
and
\be
Y_{CT}^{(2)}=-0.4981(45)\pm0.2453(35)i.
\ee
Since $|Y_{CT}^{(1)}|=0.4955(9)$ and $|Y_{CT}^{(2)}|=0.5552(42)$,
these CT singularities lie closer to the origin
than the estimated values of the FM critical point shown in Table II.
The CT critical exponents for the simple-cubic three-state Potts model
are same at $Y_{CT}^{(1)}$ and $Y_{CT}^{(2)}$ as shown in Table III,
and satisfy the scaling law $\alpha_{CT}+2\beta_{CT}+\gamma_{CT}\approx2$.
Because in three dimensions the values of the CT critical exponents for
the three-state Potts model are very close to those for the Ising model,
it is difficult not to suggest that the values of the CT critical exponents
are independent of $Q$ in three dimensions. That is, our results support the
universality of the CT singularities in three dimensions\cite{itzykson}.
For the two-dimensional $Q$-state Potts model on the square lattice
it has been shown\cite{matveev} that two pairs of the CT singularities of the model
are mapped each other through the dual transformation
\be
Y\to{1-Y\over1+(Q-1)Y},
\ee
which determines the FM critical temperature
\be
Y_c^{FM}(Q)={1\over1+\sqrt{Q}}.
\ee
Therefore, one expects the true expression for the FM critical points
of the simple-cubic Potts model to provide the mapping between
$Y_{CT}^{(1)}$ and $Y_{CT}^{(2)}$.

%%%%%%%%%%%%%%%%%%%%%%%%%%%%%%%%%%%%%%%%%%%%%%%%%%%%%%%%%%%%%%%%%%%%%%%%%%

\section{partition function zeros for an external magnetic field}

In this section we discuss the properties of the simple-cubic $Q$-state Potts models
for nonzero magnetic field using the high-field, low-temperature series
expansions\cite{sykes65,sykes73,bhanot92,enting,straley} and the zeros
of the exact partition function $Z_Q(Y,X)$
on a finite lattice in the complex temperature ($Y=e^{-\beta J}$) or
magnetic-field ($X=e^{\beta H_q}$) plane.
Figure 3 shows the partition function zeros in the complex $X$ plane of
the three-state Potts model on the $3\times 3\times 3$ simple-cubic lattice
with periodic boundary conditions in the $x$-$y$ plane
and free boundary conditions in the $z$ direction.
The zeros lie close to the unit circle at lower temperatures,
but they move away from the positive real axis and the unit circle
as the temperature is increased.
The zeros at $T=0$ are uniformly distributed on the circle with radius
$(Q-1)^{1/N_s}$ which approaches unity in the thermodynamic limit,
independent of $Q$, whereas the zeros at $T=\infty$ are $N_s$-degenerate at $X=1-Q$,
independent of lattice size\cite{kim98a,kim99,kim00a}.

The edge zero, which we call the first zero, and its complex conjugate of
a circular distribution of zeros in the complex $X$ plane cut the positive real axis
at the physical critical point $X_c=1$ for $T\le T_c$ in the thermodynamic limit.
However, for $T>T_c$ the first zero does not cut the positive real axis
in the thermodynamic limit, that is, there is a gap in the distribution of zeros
around the positive real axis. Within this gap, the free energy is analytic
and there is no phase transition.
Kortman and Griffiths\cite{kortman} carried out the first systematic investigation
of the magnetization at the first zero, based on the high-field, high-temperature
series expansion for the Ising model on the square lattice and
the diamond lattice. They found that above $T_c$ the magnetization at the first zero
diverges for the square lattice and is singular for the diamond lattice.
For $T>T_c$ we rename the first zero as the Yang-Lee edge singularity.
The divergence of the magnetization at the Yang-Lee edge singularity
means the divergence of the density of zeros, which does not occur
at a physical critical point.
Fisher\cite{fisher78} proposed the idea that the Yang-Lee edge singularity
can be thought of as a new second-order phase transition with associated critical
exponents and the Yang-Lee edge singularity can be considered
as a conventional critical point.
The critical point of the Yang-Lee edge singularity
is associated with a $\phi^3$ theory, different from the usual
critical point associated with the $\phi^4$ theory.
The crossover dimension of the Yang-Lee edge singularity is $d_c=6$.
The edge critical exponent $\sigma_e=1/\delta_e=(d-2+\eta_e)/(d+2-\eta_e)$
at a Yang-Lee edge singularity $X_e$ for $Y>Y_c$ is defined by
\be
m_e^{X}\sim\biggl(1-{X\over X_e}\biggr)^{\sigma_e},
\ee
where $m_e$ is the singular part of the magnetization in the complex
$X$ plane.
The study of the Yang-Lee edge singularity has been extended to
the classical $n$-vector model\cite{kurtze1},
the quantum Heisenberg model\cite{kurtze1},
the spherical model\cite{kurtze2},
the quantum one-dimensional transverse Ising model\cite{uzelac},
the hierarchical model\cite{baker},
the one-dimensional Potts model\cite{mittag84,glumac},
branched polymers\cite{parisi},
fluid models with repulsive-core interactions\cite{lai,ypark}, etc.
Dhar\cite{dhar} calculated the edge critical exponent $\sigma_e$ in two dimensions
by solving a particular model of three-dimensional directed animals and mapping
the solution to the hard hexagon model.
Using Fisher's idea and conformal field theory,
Cardy \cite{cardy} studied the Yang-Lee edge singularity
for a two-dimensional $\phi^3$ theory. It is generally accepted that
the value of the edge critical exponent $\sigma_e$ depends only on dimension.
The values of $\sigma_e$ are known to be $\sigma_e=-{1\over2}$ in one dimension
and $\sigma_e=-{1\over6}$\cite{dhar,cardy} in two dimensions,
whereas $\sigma_e\approx0.08$\cite{lai} in three dimensions.

The partition function zeros in the complex $X$ plane for real $Y$ have been
extensively studied and well understood. However, the partition function zeros
in the complex $Y$ plane for real $X$ are much less well understood than the
zeros in the complex $X$ plane.
Figure 4 shows the partition function zeros in the complex
$Y$ plane of the three-state Potts model for an external magnetic
field $X\ge1$ ($H_q\ge0$) on the $3\times 3\times 3$ simple-cubic lattice
with periodic boundary conditions in the $x$-$y$ plane
and free boundary conditions in the $z$ direction.
For $X=1$ some zeros distribute along the one-dimensional locus near $Y_c$,
and the first zero of the distribution cuts the positive real axis at $Y_c$ in
the thermodynamic limit as explained in the previous section.
As $H_q$ increases, all the zeros move away from the origin.
In the limit $H_q\to\infty$ ($X\to\infty$) the field $H_q$ favors the state $q$
for every site and the $Q$-state Potts model is transformed into the one-state
model whose the partition function zeros are the roots of
the equation $Y^{-N_b}=0$\cite{kim98b}.
That is, $|Y|$ for all the zeros increases without bound as $X$ increases.
Note that for $X>1$ there is accumulation of the zeros as we approach
the first zero. Matveev and Shrock\cite{matveev96a} studied the properties
of the thermodynamic functions at the first zero in the complex temperature plane
for the square-lattice Ising model with $X>1$. They observed that the thermodynamic
functions and the density of zeros diverge at the first zero when $X>1$,
and that the values of the critical exponents associated with the first zero
for $X>1$ are nearly independent of $X$ and satisfy the Rushbrooke scaling law
approximately. We call the first zero in the complex temperature plane
for $X>1$ as the Fisher edge singularity.
The edge critical exponents $\alpha_e$, $\beta_e$, and $\gamma_e$ at
the Fisher edge singularities are given by
\be
C_e\sim\biggl(1-{Y\over Y_e}\biggr)^{-\alpha_e},
\ee
\be
m_e^Y\sim\biggl(1-{Y\over Y_e}\biggr)^{\beta_e},
\ee
and
\be
\chi_e\sim\biggl(1-{Y\over Y_e}\biggr)^{-\gamma_e},
\ee
where $Y_e$ is the location of a Fisher edge singularity, and
$C_e$, $m_e^Y$ and $\chi_e$ are the singular parts of the specific heat,
magnetization, and susceptibility, respectively, for $X>1$.
Kim and Creswick\cite{kim98b} extended the study of the
Fisher edge singularity to the $Q$-state Potts model ($Q\ge3$) on the square lattice.
They found that the values of the edge critical exponents at the Fisher edge singularities
are nearly independent of $Q$ ($Q\ge2$).

In three dimensions to study the critical behavior at the Fisher edge singularity
we have used the high-field, low-temperature series expansions
for the Ising model\cite{sykes65,sykes73,bhanot92} and
the three-state Potts model\cite{enting,straley} on the simple-cubic lattice.
Table IV shows the locations ($Y_e$) and the edge critical exponents
($\alpha_e$, $\beta_e$, and $\gamma_e$)
of the Fisher edge singularities estimated by Dlog Pad\'e approximants
to the specific heat, magnetization, and susceptibility series
for $x=100$, 200, and 500. The edge critical exponents, which
are independent of $X$, have the values of $\alpha_e=1.0\sim1.1$, $\beta_e\approx-0.4$,
and $\gamma_e\approx1.0$ for $Q=2$ and for $Q=3$ similar values of
$\alpha_e$ and $\gamma_e$ and $\beta_e\approx-0.6$. So in three dimensions
we seem to have a strong violation of the scaling law
$\alpha_e+2\beta_e+\gamma_e=2$.
Another interesting observation is that there exist apparently three pairs
of the Fisher edge singularities in three dimensions (Figure 4 and Table IV).
The estimates of the edge critical exponents at each of the three edge
singularities indicate that they are equal within error bars.

On the other hand, everything can be evaluated exactly in one dimension,
and there exist only one pair
of the Fisher edge singularities characterized by the edge critical exponents
$\alpha_e={3\over2}$, $\beta_e=-{1\over2}$, and $\gamma_e={3\over2}$,
all independent of $Q$. These values satisfy the scaling law
$\alpha_e+2\beta_e+\gamma_e=2$.
Table V shows the locations and the edge critical exponents of the Fisher
edge singularities for the $Q$-state Potts models on the square lattice.
In two dimensions we have also found two pairs of the Fisher edge singularities
which have the same values of the edge critical exponents within error bars.
The edge critical exponents have the values of $\alpha_e=1.13\sim1.24$,
$\beta_e=-0.11\sim-0.20$, and $\gamma_e=1.11\sim1.26$.
In two dimensions the edge critical exponents also satisfy
the scaling law $\alpha_e+2\beta_e+\gamma_e=2$. Now we can conclude that
the number of pairs of the Fisher edge singularities is equal to the dimensionality.

Until now the Yang-Lee and Fisher edge singularities have been studied
and understood independently. However, the Yang-Lee edge singularity seems to have
a direct relationship to the Fisher edge singularity.
From the edge critical exponents of the Fisher edge singularity in one dimension
we get $y_t^e=2$ and $y_h^e=2$, which also follows from
$\sigma_e=-{1\over2}$ at the Yang-Lee edge singularity. The equivalence
of the Yang-Lee and Fisher edge singularities is not surprising
in light of the transformation which maps the temperature and field
variables into each other in one dimension\cite{glumac}.
In two dimensions it is known that $\sigma_e=-{1\over6}$ from which it
follows that $y_h^e={12\over5}$.
If we assume $y_t^e=y_h^e={12\over5}$ for the Fisher edge
singularity, we have $\alpha_e={7\over6}$, $\beta_e=-{1\over6}$, and
$\gamma_e={7\over6}$, which are not far from the values of the exponents
estimated by series expansions.
In three dimensions if we take $\alpha_e=1.0$ then $y_t^e=3$, and $\sigma_e\sim0$
(logarithmic singularity), which seems to be the historical trend,
then $y_h^e=3$ as well. So in $d=1$, 2 and 3 we seem to find $y_t^e\sim y_h^e$.
If $y_t^e=y_h^e$, the Fisher and Yang-Lee edge singularities are not independent
behaviors, and they seem to have only one relevant variable.

%%%%%%%%%%%%%%%%%%%%%%%%%%%%%%%%%%%%%%%%%%%%%%%%%%%%%%%%%%%%%%%%%%%%%%%%%%%

\section{conclusion}

We have investigated the interesting properties of the $Q$-state Potts model
on the simple-cubic lattice using the zeros of the exact partition function
on a finite lattice and series expansions.
The critical behavior of the Potts model in the ferromagnetic and
antiferromagnetic phases has been studied at the same time based on the distribution
of the partition function zeros in the complex temperature plane.
The distribution of the zeros in the complex temperature plane reveals
the possibility of the intermediate phase transitions below the order-disorder
phase transition of the Potts antiferromagnet.
The behavior at the CT singularities, where some thermodynamic quantities diverge
with the characteristic CT exponents $\alpha_{CT}$, $\beta_{CT}$, and $\gamma_{CT}$,
has been discussed using the partition function zeros and the low-temperature series
expansion of the Potts model in the absence of an external magnetic field.
In the complex temperature plane the CT singularities coexist with the physical
critical points for $H_q=0$.

We have also studied the veiled properties of the Potts model for nonzero magnetic field
using the partition function zeros and the high-field, low-temperature series expansion.
As $H_q$ increases, all the zeros in the complex $Y=e^{\beta J}$ plane move away from
the origin. For $H_q>0$ the physical critical points $Y_c^{FM}$ and $Y_c^{AFM}$
disappear and the Fisher edge singularities $Y_e$ appear in the complex temperature plane.
For different values of $X=e^{\beta H_q}$
we have calculated the locations and the edge critical exponents of the Fisher
edge singularities of the Ising and three-state Potts models
on the simple-cubic lattice.
The values of the edge critical exponents $\alpha_e$, $\beta_e$, and $\gamma_e$
at a Fisher edge singularity are universal, that is, independent of $X$ and $Q$
but dependent on dimension $d$.
We found that the number of pairs of the Fisher edge singularities is equal to $d$.
We have considered the Yang-Lee edge singularity $X_e$, which appears in the
complex $X$ plane for $Y>Y_c$ instead of the physical critical point $X_c=1$,
and its edge exponent $\sigma_e=1/\delta_e$.
In one, two, and three dimensions the Yang-Lee edge singularity is related to
the Fisher edge singularity. They seem to have only one relevant variable
$y_t^e\sim y_h^e$.

%%%%%%%%%%%%%%%%%%%%%%%%%%%%%%%%%%%%%%%%%%%%%%%%%%%%%%%%%%%%%%%%%%%%%%%%%%

\vskip2cm
\begin{center}
ACKNOWLEDGMENTS
\end{center}

The author is indebted to Prof. R. J. Creswick for very helpful discussions.
The author is grateful to Prof. A. J. Guttmann for kindly providing him with
the missed coefficients in the paper\cite{guttmann1}.
The author also wishes to thank Prof. M. E. Fisher for valuable comments
and for kindly drawing his attention to the paper\cite{ypark}.

%%%%%%%%%%%%%%%%%%%%%%%%%%%%%%%%%%%%%%%%%%%%%%%%%%%%%%%%%%%%%%%%%%%%%%%%%%%

%%%%%%%%%%%%%%%%%%%%%%%%%%%%%%%%%%%%%%%%%%%%%%%%%%%%%%%%%%%%%%%%%%%%%%%%%%

\newpage

\begin{table}
\caption{The density of states $\Omega_{Q=7}(E)$ of
the seven-state Potts model on the $3\times3\times3$ simple-cubic lattice.}
\begin{ruledtabular}
\begin{tabular}{cr}
$E$ &$\Omega_{Q=7}(E)$ \ \ \ \ \ \ \ \ \ \ \ \\
\hline
0  &7 \\
1  &0 \\
2  &0 \\
3  &0 \\
4  &0 \\
5  &756 \\
6  &378 \\
$\vdots$ &$\vdots$ \\
61 &8297099724190402024200 \\
62 &8931912475351510844640 \\
63 &8562381313020489303720 \\
$\vdots$ &$\vdots$ \\
70 &60500919374783049960 \\
71 &9822183432973269120 \\
72 &794516261799074640 \\
\end{tabular}
\end{ruledtabular}
\end{table}

\begin{table}
\caption{The ferromagnetic critical temperatures $Y_c^{FM}$ of the $Q$-state Potts
models on the simple-cubic lattice. $Y_c^{FM}$(numer.) are the values of the critical
temperatures estimated by various numerical methods in the literature.}
\begin{ruledtabular}
\begin{tabular}{cccl}
$Q$ &$Y_c^{FM}=1/(Q+\sqrt{2Q-1})^{1/3}$
&$Y_c^{FM}=1/(\pi^{-1}+Q^{1/3})$ &$Y_c^{FM}$(numer.)\\
\hline
1 &0.793700... &0.758546... &0.753\cite{gaunt,stauffer} \\
2 &0.644689... &0.633620... &0.6419\cite{landau,ballesteros} \\
3 &0.575879... &0.568001... &0.571\cite{dkim} \\
  &            &            &0.5766\cite{janke} \\
  &            &            &0.5767\cite{guttmann1} \\
4 &0.531886... &0.524738... &0.523\cite{dkim} \\
  &            &            &0.524\cite{ditzian} \\
  &            &            &0.532\cite{park} \\
5 &$1\over2$   &0.493027... &0.501\cite{park} \\
6 &0.475240... &0.468289... &0.472\cite{wu} \\
  &            &            &0.477\cite{park} \\
7 &0.455151... &0.448181... &0.456\cite{chen} \\
\end{tabular}
\end{ruledtabular}
\end{table}

\begin{table}
\caption{The locations ($A_{CT}=Y_{CT}^{-1}$) and the corresponding critical exponents
of the complex-temperature (CT) singularities for the three-state Potts models
on the simple-cubic lattice.}
\begin{ruledtabular}
\begin{tabular}{cllll}
$A_{CT}$ &\quad $\alpha_{CT}$ &\qquad $\beta_{CT}$ &\quad $\gamma_{CT}$ &$\alpha$+2$\beta$+$\gamma$ \\
\hline
 $0.1036(49) \pm2.0157(35)i$  &1.107(8)   &$-0.063(1)$ &1.161(38)  &2.142(39)  \\
$-1.6158(119)\pm0.7958(142)i$ &1.208(212) &$-0.063(2)$ &1.237(166) &2.319(269) \\
\end{tabular}
\end{ruledtabular}
\end{table}

\begin{table}
\caption{The locations and the edge critical exponents of the Fisher edge singularities
for the Ising ($Q=2$) and three-state Potts ($Q=3$) models
on the simple-cubic lattice for an external magnetic field $X=e^{\beta H_q}$.}
\begin{ruledtabular}
\begin{tabular}{ccllll}
$X$ &$Y_e$ &\quad $\alpha_e$ &\qquad $\beta_e$ &\quad $\gamma_e$ &$\alpha_e$+2$\beta_e$+$\gamma_e$ \\
\hline
100($Q=2$)  &$1.249(9) \pm0.642(4)i$  &1.079(54)  &$-0.418(27)$ &1.024(10)  &1.267(67) \\
                     &$\pm1.323(12)i$ &1.047(64)  &$-0.410(86)$ &1.075(78)  &1.303(157) \\
            &$-1.249(9)\pm0.642(4)i$  &1.079(54)  &$-0.418(27)$ &1.024(10)  &1.267(67) \\
200($Q=2$)  &$1.433(10)\pm0.737(4)i$  &1.046(62)  &$-0.422(24)$ &1.031(9)   &1.233(71) \\
                     &$\pm1.500(12)i$ &1.052(2)   &$-0.415(107)$&1.058(125) &1.280(197) \\
           &$-1.433(10)\pm0.737(4)i$  &1.046(62)  &$-0.422(24)$ &1.031(9)   &1.233(71) \\
500($Q=2$)  &$1.649(11)\pm0.876(5)i$  &1.057(24)  &$-0.425(24)$ &1.038(9)   &1.246(42) \\
                     &$\pm1.771(21)i$ &1.050(5)   &$-0.418(76)$ &1.058(114) &1.271(157) \\
           &$-1.649(11)\pm0.876(5)i$  &1.057(24)  &$-0.425(24)$ &1.038(9)   &1.246(42) \\
100($Q=3$)  &$1.217(35)\pm0.573(21)i$ &1.065(141) &$-0.611(50)$ &1.003(33)  &0.846(161) \\
            &$0.052(13)\pm1.201(32)i$ &0.999(155) &$-0.608(14)$ &1.076(134) &0.858(206) \\
           &$-1.070(25)\pm0.603(17)i$ &1.070(158) &$-0.603(17)$ &1.064(60)  &0.927(171) \\
200($Q=3$)  &$1.342(36)\pm0.660(22)i$ &1.075(35)  &$-0.605(16)$ &0.998(82)  &0.863(92) \\
            &$0.054(11)\pm1.359(38)i$ &0.996(245) &$-0.610(26)$ &1.064(137) &0.839(283) \\
           &$-1.199(29)\pm0.682(21)i$ &1.039(227) &$-0.610(20)$ &1.066(66)  &0.885(238) \\
500($Q=3$)  &$1.535(39)\pm0.787(24)i$ &1.071(17)  &$-0.605(48)$ &1.012(22)  &0.872(73) \\
             &$0.056(8)\pm1.601(41)i$ &1.058(134) &$-0.615(32)$ &1.076(106) &0.905(177) \\
           &$-1.397(33)\pm0.802(22)i$ &1.096(66)  &$-0.613(37)$ &1.063(155) &0.934(177) \\
\end{tabular}
\end{ruledtabular}
\end{table}

\begin{table}
\caption{The locations and the edge critical exponents of the Fisher edge singularities
for the $Q$-state Potts models on the square lattice for $X=100$.}
\begin{ruledtabular}
\begin{tabular}{ccllll}
$Q$ &$Y_e$ &\quad $\alpha_e$ &\qquad $\beta_e$ &\quad $\gamma_e$ &$\alpha_e$+2$\beta_e$+$\gamma_e$ \\
\hline
3   &$1.232(2)\pm1.048(3)i$  &1.217(19)  &$-0.197(6)$  &1.205(33)  &2.029(39) \\
  &$-1.063(13)\pm1.068(6)i$  &1.238(26)  &$-0.172(25)$ &1.257(69)  &2.151(82) \\
4   &$1.159(5)\pm0.929(10)i$ &1.181(76)  &$-0.180(4)$  &1.121(77)  &1.941(76) \\
  &$-0.908(13)\pm0.959(10)i$ &1.136(17)  &$-0.127(11)$ &1.206(16)  &2.088(28) \\
5   &$1.103(4)\pm0.859(14)i$ &1.181(109) &$-0.173(4)$  &1.110(98)  &1.946(109) \\
  &$-0.834(18)\pm0.889(27)i$ &1.131(176) &$-0.109(28)$ &1.207(166) &2.120(245) \\
\end{tabular}
\end{ruledtabular}
\end{table}

%%%%%%%%%%%%%%%%%%%%%%%%%%%%%%%%%%%%%%%%%%%%%%%%%%%%%%%%%%%%%%%%%%%%%%%%%%

\begin{figure}
\epsfbox{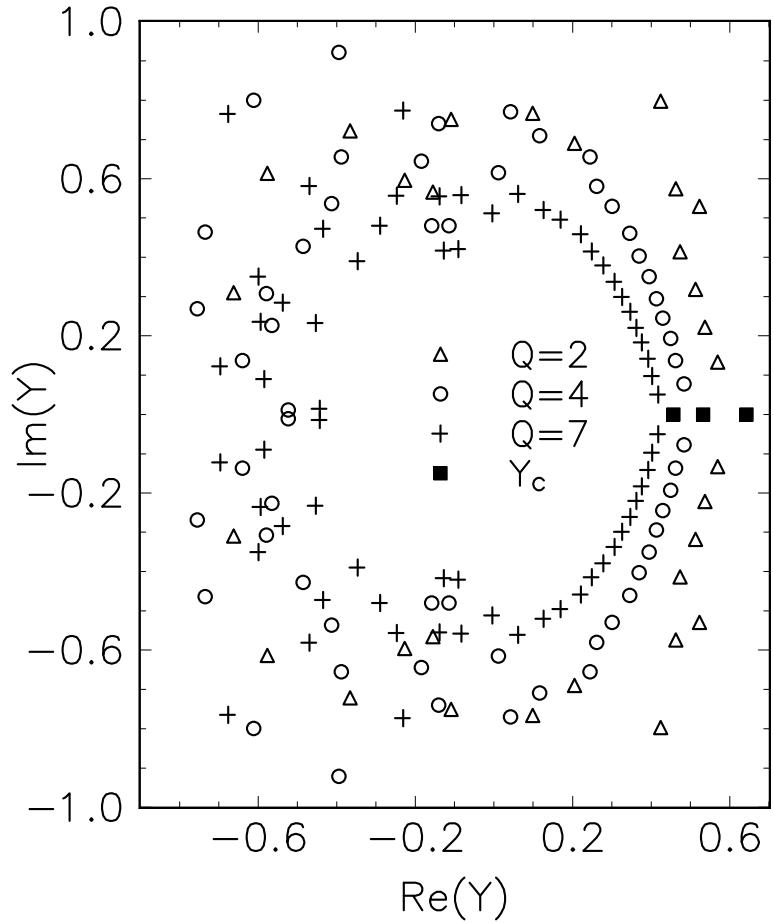}
\caption{Partition function zeros in the complex
$Y=e^{-\beta J}$ plane of the $Q$-state Potts models on the
$3\times 3\times 3$ simple-cubic lattice. The solid squares
($Y_c$) show the locations of the ferromagnetic critical points
estimated in the literature for $Q=2$\cite{landau,ballesteros},
4\cite{park}, and 7\cite{chen}.}
\end{figure}

\begin{figure}
\epsfbox{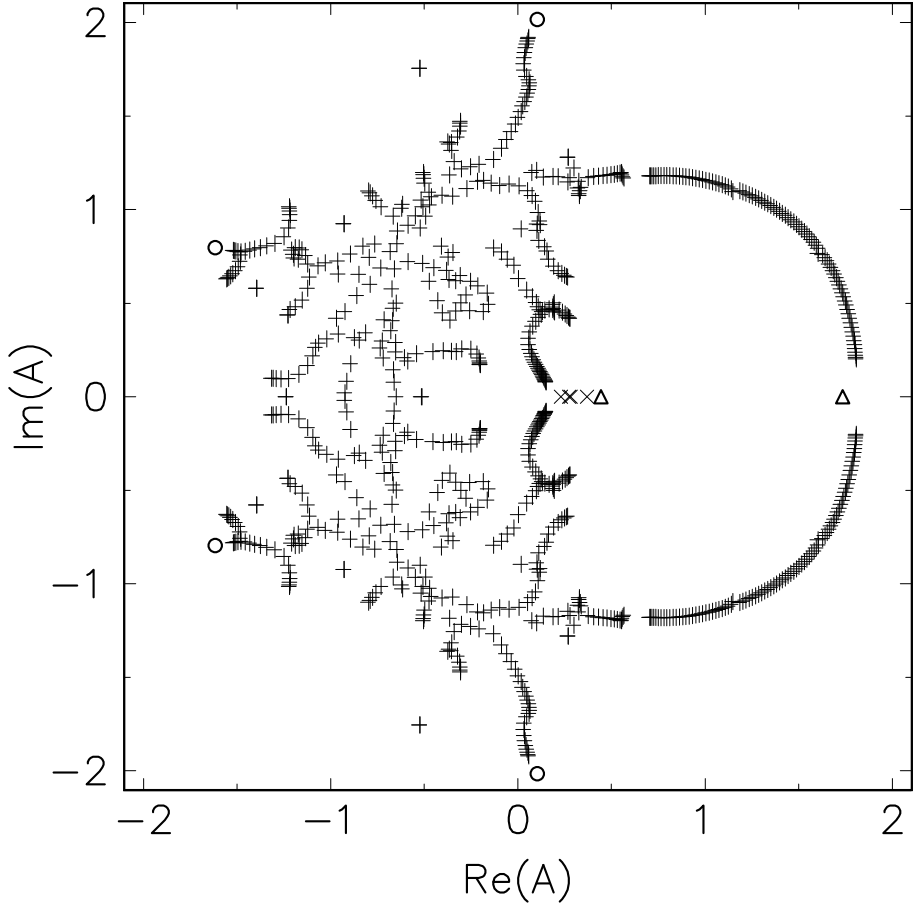}
\caption{Partition function zeros in the complex
$A=e^{\beta J}$ plane of the three-state Potts model on the
$3\times 3\times 30$ simple-cubic lattice. The open triangles
show the locations of the ferromagnetic ($A_c^{FM}=1.7342$)\cite{janke}
and antiferromagnetic ($A_c^{AFM}=0.4424$)\cite{wang} critical points
estimated in the literature.
The X marks are the suggested locations of additional transition temperatures
at $A_{KH}=0.3679$\cite{kolesik,heilmann},
$A_{RL1}=0.2806$ and $A_{RL2}=0.2736$\cite{rosengren93} (Because $A_{RL1}$
and $A_{RL2}$ are too close, they are not clearly distinguished in the Figure.),
and $A_{KRL}=0.2298$\cite{kundrotas} below $A_c^{AFM}$
for the antiferromagnetic model.
The open circles are the locations of two pairs of the complex-temperature (CT)
singularities estimated from the series analysis.}
\end{figure}

\begin{figure}
\epsfbox{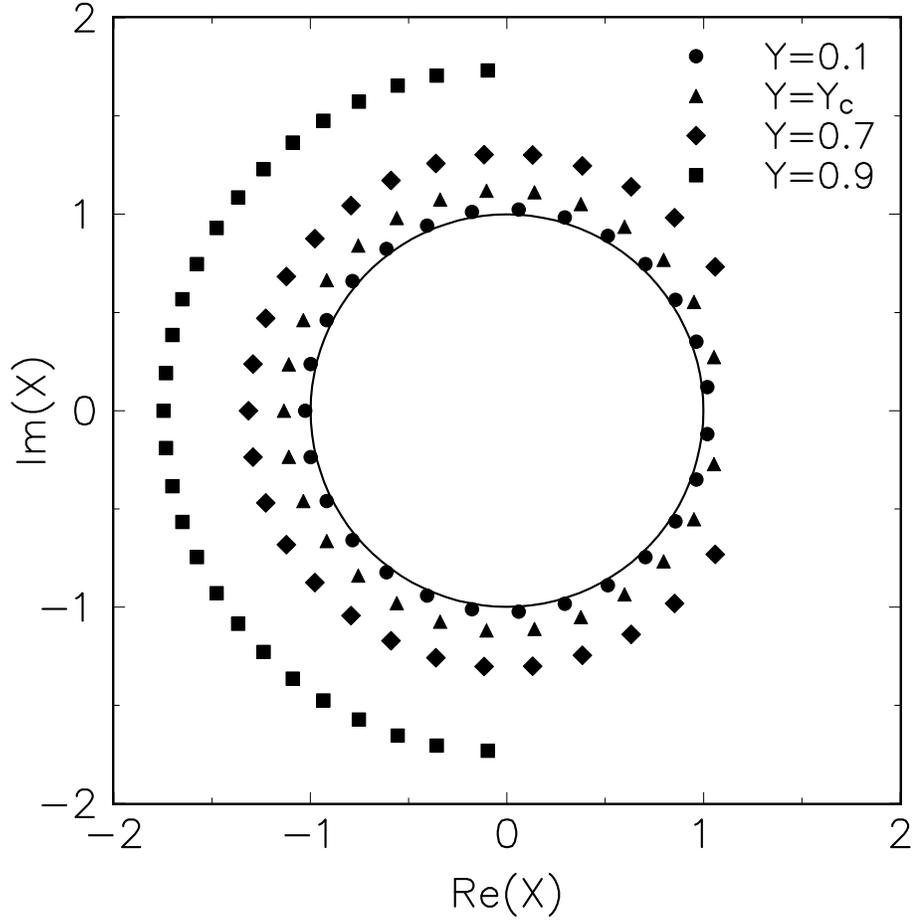}
\caption{Partition function zeros in the complex
$X=e^{\beta H_q}$ plane of the three-state Potts model on the
$3\times 3\times 3$ simple-cubic lattice for $Y=e^{-\beta J}=0.1$,
$Y=Y_c\approx0.5766$\cite{janke}, $Y=0.7$, and $Y=0.9$.}
\end{figure}

\begin{figure}
\epsfbox{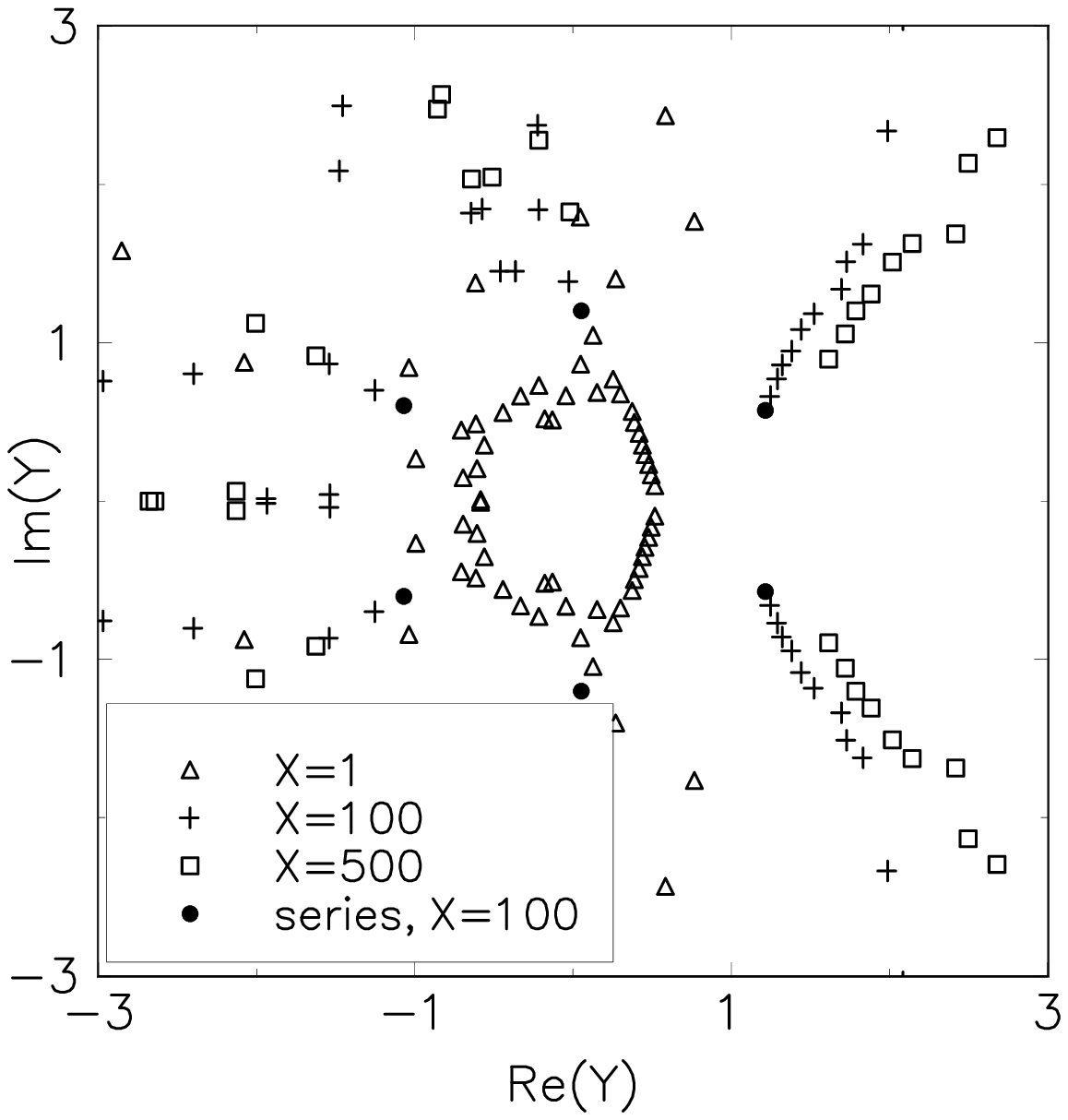}
\caption{Partition function zeros in the complex
$Y=e^{-\beta J}$ plane of the three-state Potts model on the
$3\times 3\times 3$ simple-cubic lattice for an external magnetic
field $X=e^{\beta H_q}$. The closed circles show the locations of
three pairs of the Fisher edge singularities estimated from the
series analysis for $X=100$.}
\end{figure}

%%%%%%%%%%%%%%%%%%%%%%%%%%%%%%%%%%%%%%%%%%%%%%%%%%%%%%%%%%%%%%%%%%%%%%%%%%


\newpage

\begin{thebibliography}{199}

\bibitem{wu} F. Y. Wu, Rev. Mod. Phys. 54 (1982) 235.
\bibitem{baxter73} R. J. Baxter, J. Phys. C 6 (1973) L445.
\bibitem{brown} F. R. Brown, N. H. Christ, Y. Deng, M. Gao, and T. J. Woch,
Phys. Rev. Lett. 61 (1988) 2058.
\bibitem{alves90a} N. A. Alves, B. A. Berg, and S. Sanielevici,
Phys. Rev. Lett. 64 (1990) 3107.
\bibitem{kolb} E. W. Kolb and M. S. Turner, The Early Universe
(Addison-Wesley, Reading, 1994).
\bibitem{jlee} J. Lee and J. M. Kosterlitz, Phys. Rev. B 43 (1991) 1286.
\bibitem{svetitsky} B. Svetitsky and L. G. Yaffe,
Nucl. Phys. B 210 (1982) 423.
\bibitem{baxter82} R. J. Baxter, Proc. R. Soc. London A 383 (1982) 43.
\bibitem{banavar} J. R. Banavar, G. S. Grest, and D. Jasnow,
Phys. Rev. B 25 (1982) 4639.
\bibitem{wang} J.-S. Wang, R. H. Swendsen, and R. Koteck\'{y},
Phys. Rev. B 42 (1990) 2465.
\bibitem{rosengren93} A. Rosengren and S. Lapinskas,
Phys. Rev. Lett. 71 (1993) 165.
\bibitem{kundrotas} P. J. Kundrotas, A. Rosengren and S. Lapinskas,
Phys. Rev. B 52 (1995) 9166.
\bibitem{kolesik} M. Kolesik and M. Suzuki, J. Phys. A 28 (1995) 6543.
\bibitem{heilmann} R. K. Heilmann, J.-S. Wang, and R. H. Swendsen,
Phys. Rev. B 53 (1996) 2210.
\bibitem{oshikawa} M. Oshikawa, Phys. Rev. B 61 (2000) 3430.
\bibitem{yang} C. N. Yang and T. D. Lee, Phys. Rev. 87 (1952) 404.
\bibitem{tdlee} T. D. Lee and C. N. Yang, Phys. Rev. 87 (1952) 410.
\bibitem{creswick97} R. J. Creswick and S.-Y. Kim,
Phys. Rev. E 56 (1997) 2418.
\bibitem{creswick98}
R. J. Creswick and S.-Y. Kim, in: D. P. Landau, K. K. Mon,
and H.-B. Sch\"{u}ttler (Ed.), Computer Simulation Studies
in Condensed-Matter Physics, Vol. 10, Springer, Berlin, 1998, p. 224.
\bibitem{creswick99} R. J. Creswick and S.-Y. Kim,
Comput. Phys. Commun. 121 (1999) 26.
\bibitem{janke01} W. Janke and R. Kenna,
J. Stat. Phys. 102 (2001) 1211.
\bibitem{suzuki} M. Suzuki, C. Kawabata, S. Ono, Y. Karaki, and M. Ikeda,
J. Phys. Soc. Japan 29 (1970) 837.
\bibitem{kortman} P. J. Kortman and R. B. Griffiths,
Phys. Rev. Lett. 27 (1971) 1439.
\bibitem{kim98a} S.-Y. Kim and R. J. Creswick,
Phys. Rev. Lett. 81 (1998) 2000.
\bibitem{kim99} S.-Y. Kim and R. J. Creswick,
Phys. Rev. Lett. 82 (1999) 3924.
\bibitem{kim00a} S.-Y. Kim and R. J. Creswick,
Physica A 281 (2000) 252.
\bibitem{fisher65} M. E. Fisher,
in: W. E. Brittin (Ed.), Lectures in Theoretical Physics, Vol. 7c,
University of Colorado Press, Boulder, 1965, p. 1.
\bibitem{ono} S. Ono, Y. Karaki, M. Suzuki, and C. Kawabata,
J. Phys. Soc. Japan 25 (1968) 54.
\bibitem{pearson} R. B. Pearson, Phys. Rev. B 26 (1982) 6285.
\bibitem{itzykson} C. Itzykson, R. B. Pearson, and J. B. Zuber,
Nucl. Phys. B 220 (1983) 415.
\bibitem{martin83a} P. P. Martin, Nucl. Phys. B 220 (1983) 366.
\bibitem{bhanot90a} G. Bhanot and S. Sastry,
J. Stat. Phys. 60 (1990) 333.
\bibitem{marinari} E. Marinari, Nucl. Phys. B 235 (1984) 123.
\bibitem{bhanot87} G. Bhanot, R. Salvador, S. Black, P. Carter, and R. Toral,
Phys. Rev. Lett. 59 (1987) 803.
\bibitem{alves90b} N. A. Alves, B. A. Berg, and R. Villanova,
Phys. Rev. B 41 (1990) 383.
\bibitem{alves00} N. A. Alves, J. R. Drugowich de Felicio, and U. H. E. Hansmann,
J. Phys. A 33 (2000) 7489.
\bibitem{martin86} P. P. Martin, J. Phys. A 19 (1986) 3267.
\bibitem{cnchen} C.-N. Chen, C.-K. Hu, and F. Y. Wu,
Phys. Rev. Lett. 76 (1996) 169.
\bibitem{kim98b} S.-Y. Kim and R. J. Creswick,
Phys. Rev. E 58 (1998) 7006.
\bibitem{kim00b} S.-Y. Kim, R. J. Creswick, C.-N. Chen, and C.-K. Hu,
Physica A 281 (2000) 262.
\bibitem{kim01} S.-Y. Kim and R. J. Creswick,
Phys. Rev. E 63 (2001) 066107.
\bibitem{martin83b} P. P. Martin, Nucl. Phys. B 225 (1983) 497.
\bibitem{alves91} N. A. Alves, B. A. Berg, and R. Villanova,
Phys. Rev. B 43 (1991) 5846.
\bibitem{bhanot90b} G. Bhanot, J. Stat. Phys. 60 (1990) 55.
\bibitem{stosic} B. Sto\v{s}i\'c, S. Milo\v{s}evi\'c, and H. E. Stanley,
Phys. Rev. B 41 (1990) 11466.
\bibitem{stodolsky} L. Stodolsky and J. Wosiek,
Nucl. Phys. B 413 (1994) 813.
\bibitem{creswick95} R. J. Creswick, Phys. Rev. E 52 (1995) 5735.
\bibitem{mittag} L. Mittag and M. J. Stephen,
J. Phys. A 7 (1974) L109.
\bibitem{pearce} P. A. Pearce and R. B. Griffiths,
J. Phys. A 13 (1980) 2143.
\bibitem{rosengren1} A. Rosengren, J. Phys. A 19 (1986) 1709.
\bibitem{hajdukovic} D. Hajdukovi\'c, J. Phys. A 16 (1983) L193.
\bibitem{gaunt} D. S. Gaunt and H. Ruskin, J. Phys. A 11 (1978) 1369.
\bibitem{stauffer} D. Stauffer, Phys. Rep. 54 (1979) 1.
\bibitem{landau} D. P. Landau, Physica A 205 (1994) 41.
\bibitem{ballesteros} H. G. Ballesteros, L. A. Fernandez, V. Martin-Mayor,
A. Munoz Sudupe, G. Parisi, and J. J. Ruiz-Lorenzo,
J. Phys. A 32 (1999) 1.
\bibitem{dkim} D. Kim and R. I. Joseph, J. Phys. A 8 (1975) 891.
\bibitem{janke} W. Janke and R. Villanova,
Nucl. Phys. B 489 (1997) 679.
\bibitem{guttmann1} A. J. Guttmann and I. G. Enting,
J. Phys. A 27 (1994) 5801.
This paper has the longest low-temperature series for the specific heat and
susceptibility of the simple-cubic three-state Potts model.
The coefficients of $(Y^2)^{36}$ were unintentionally omitted by publisher.
The missed coefficients are $5858732$, $-58983558$, and $281636092$
for $\lambda_{36}$, $m_{36}$, and $c_{36}$, respectively.
And the coefficient $c_{21}=36922$ is a misprint. It should be $-36922$.
\bibitem{ditzian} R. V. Ditzian and L. P. Kadanoff, J. Phys. A 12 (1979) L229.
\bibitem{park} H. Park and D. Kim, J. Korean Phys. Soc. 15 (1982) 55.
\bibitem{chen} H. H. Chen, F. Lee, and Y. M. Kao,
Phys. Rev. B 52 (1995) 39.
\bibitem{vohwinkel} C. Vohwinkel, Phys. Lett. B 301 (1993) 208.
This paper has the low-temperature series only for the spontaneous magnetization
of the simple-cubic three-state Potts model.
But it is the longest series for the spontaneous magnetization.
\bibitem{bhanot} G. Bhanot, M. Creutz, U. Gl\"{a}ssner, I. Horvath, J. Lacki,
K. Schilling, and J. Weckel, Phys. Rev. B 48 (1993) 6183.
In this paper, the coefficient $c_{29}=12285816$ for the susceptibility series
is a misprint, instead, $-12285816$ is correct.
\bibitem{guttmann2} C. J. Thompson, A. J. Guttmann, and B. W. Ninham,
J. Phys. C 2 (1969) 1889.
\bibitem{guttmann3} A. J. Guttmann, J. Phys. C 2 (1969) 1900.
\bibitem{guttmann4} C. Domb and A. J. Guttmann,
J. Phys. C 3 (1970) 1652.
\bibitem{guttmann5} A. J. Guttmann and I. G. Enting,
J. Phys. A 26 (1993) 807.
\bibitem{guttmann6} A. J. Guttmann, in: C. Domb and J. Lebowitz (Ed.),
Phase Transitions and Critical Phenomena, Vol. 13,
Academic, New York, 1989, p. 1.
\bibitem{matveev} V. Matveev and R. Shrock,
Phys. Rev. E 54 (1996) 6174.
\bibitem{sykes65} M. F. Sykes, J. W. Essam, and D. S. Gaunt,
J. Math. Phys. 6 (1965) 283.
\bibitem{sykes73} M. F. Sykes, D. S. Gaunt, J. W. Essam, and C. J. Elliott,
J. Phys. A 6 (1973) 1507.
\bibitem{bhanot92} G. Bhanot, M. Creutz, and J. Lacki,
Phys. Rev. Lett. 69 (1992) 1841.
\bibitem{enting} I. G. Enting, J. Phys. A 7 (1974) 1617.
\bibitem{straley} J. P. Straley, J. Phys. A 7 (1974) 2173.
\bibitem{fisher78} M. E. Fisher, Phys. Rev. Lett. 40 (1978) 1610.
\bibitem{kurtze1} D. A. Kurtze and M. E. Fisher,
Phys. Rev. B 20 (1979) 2785.
\bibitem{kurtze2} D. A. Kurtze and M. E. Fisher,
J. Stat. Phys. 19 (1978) 205.
\bibitem{uzelac} K. Uzelac, P. Pfeuty, and R. Jullien,
Phys. Rev. Lett. 43 (1979) 805.
\bibitem{baker} G. A. Baker, Jr., M. E. Fisher, and P. Moussa,
Phys. Rev. Lett. 42 (1979) 615.
\bibitem{mittag84} L. Mittag and M. J. Stephen, J. Stat. Phys. 35 (1984) 303.
\bibitem{glumac} Z. Glumac and K. Uzelac, J. Phys. A 27 (1994) 7709.
\bibitem{parisi} G. Parisi and N. Sourlas,
Phys. Rev. Lett. 46 (1981) 871.
\bibitem{lai} S.-N. Lai and M. E. Fisher,
J. Chem. Phys. 103 (1995) 8144.
\bibitem{ypark} Y. Park and M. E. Fisher,
Phys. Rev. E 60 (1999) 6323.
\bibitem{dhar} D. Dhar, Phys. Rev. Lett. 51 (1983) 853.
\bibitem{cardy} J. L. Cardy, Phys. Rev. Lett. 54 (1985) 1354.
\bibitem{matveev96a} V. Matveev and R. Shrock,
Phys. Rev. E 53 (1996) 254.

\end{thebibliography}
\end{document}